\documentclass[sigconf,nonacm]{acmart}
\AtBeginDocument{%
  }

\usepackage{soul}

\begin{document}

\title[Web Crawler Restrictions, AI Training Datasets \& Political Biases]{Web Crawler Restrictions, AI Training Datasets\\ \& Political Biases}

\author{
    Paul Bouchaud$^{1,2}$,
	Pedro Ramaciotti$^{1,2,3}$\and
	\small$^{1}$Complex Systems Institute of Paris Ile-de-France CNRS, Paris, France.\and
    \small$^{2}$médialab, Sciences Po, Paris, France.\and    \small$^{3}$Learning Planet Institute, CY Cergy Paris University, Paris, France.
}

\renewcommand{\shortauthors}{Bouchaud and Ramaciotti}

\begin{abstract}

Large language models rely on web-scraped text for training; concurrently, content creators are increasingly blocking AI crawlers to retain control over their data. We analyze crawler restrictions across the top one million most-visited websites since 2023 and examine their potential downstream effects on training data composition. Our analysis reveals growing restrictions, with blocking patterns varying by website popularity and content type. A quarter of the top thousand websites restrict AI crawlers, decreasing to one-tenth across the broader top million. Content type matters significantly: 34.2\% of news outlets disallow OpenAI's \texttt{GPTBot}, rising to 55\% for outlets with high factual reporting. Additionally, outlets with neutral political positions impose the strongest restrictions (58\%), whereas hyperpartisan websites and those with low factual reporting impose fewer restrictions ---only 4.1\% of right-leaning outlets block access to OpenAI. Our findings suggest that heterogeneous blocking patterns may skew training datasets toward low-quality or polarized content, potentially affecting the capabilities of models served by prominent AI-as-a-Service providers.

\end{abstract}



\received{7 October 2025}

\maketitle

\section{Introduction}

The training of large language models that powered the recent surge in AI chatbots crucially relied on massive text corpora \cite{llama, gpt3} ---amounting to trillions of words scraped from billions of web pages \cite{commoncrawl}. This large-scale data collection, and its subsequent commercial exploitation, has raised fundamental copyright concerns and prompted content creators to resist having their work used without compensation or attribution \cite{grynbaum2023times, Lucchi2023,Strowel2023}. Following the training of their flagship model, OpenAI announced in August 2023 support for the Robots Exclusion Protocol \cite{openaiCrawlers}, a standard initiated in the late 1990s that allows webmasters to specify which web crawlers can access their content by placing directives in a \texttt{robots.txt} file at the root of their websites. Following this development, several organizations and content creators called for blocking OpenAI's and other AI providers' crawlers \cite{stubblebine2023, NEURIPS2024_c3738949, 2024arXiv241115091L} as a means to exert greater control over the use of their data. 

Meanwhile, since the emergence of natural language processing, a wealth of work has addressed how biases in training corpora might translate to biased outcomes. Gender and racial biases \cite{Caliskan2017, Rozado2020, May2019} in word embedding models such as word2vec \cite{word2vec} and GloVe \cite{pennington2014glove} and subsequent sentence-level embeddings such as BERT \cite{devlin2019bert}, were traceable to biases in the underlying training corpus \cite{bolukbasi2016, feng-etal-2023-pretraining}. However, the increased size of modern model representations and datasets, coupled with opacity regarding training data composition, impedes such investigations. Nonetheless, one resource remains central to most AI training efforts \cite{llama, gpt3, Baack2024}: CommonCrawl \cite{commoncrawl}.

CommonCrawl is a petabyte-large archive of web crawl data collected since 2008 \cite{commoncrawl, Baack2024}. Due to its size and coverage, it is massively used in the training of AI generative models, in particular through derivative datasets. Indeed, CommonCrawl collectors have taken a stated aim not to curate or polish the content of the collection, capturing a fraction of the web as it is \cite{Baack2024}. Consequently, large language model builders rely on filtered versions of CommonCrawl such as C4 \cite{c4}, RefineWeb \cite{penedo2023refinedwebdatasetfalconllm}, or FineWeb \cite{fineweb} that typically remove duplicates, HTML boilerplate, and subsample certain languages \cite{ccnet}. Some curation rules apply normative judgments by filtering out content deemed ``undesirable'' \cite{fineweb, Baack2024, luccioni-viviano-2021-whats} and have been shown to disproportionately exclude text from and about minorities \cite{confc4}, potentially skewing the representation in models trained on such filtered datasets.

In addition to content filtering, another potential source of bias may arise directly from the collection process itself, as AI crawlers face growing restrictions \cite{2024arXiv241115091L, dinzinger2024surveywebcontentcontrol}. If AI model builders do not properly address the skew induced by the protection measures content creators adopt against exploitative practices, the trained AI models may inherit systematic biases that may compromise their fairness and representativeness. In this work, we quantify and characterize the adoption of AI crawler restrictions since 2023 across the one million most visited websites worldwide, and evaluate the downstream effects on CommonCrawl and its derivative training datasets.

Our analysis reveals growing AI crawler restrictions since 2023, with blocking patterns varying by website popularity and content type. Over 25\% of the top thousand most visited websites restrict AI crawlers, but this proportion decreases to 10\% when considering the broader top million websites. Content type also plays a significant role: while only 4\% of shopping websites disallow OpenAI's \texttt{GPTBot}, 34.2\% of news outlets do, rising to 55\% for outlets with high factual reporting. Outlets with neutral political positions impose the strongest restrictions, with 58\% disallowing OpenAI's crawler, whereas hyperpartisan websites and those with low factual reporting impose fewer restrictions---only 4.1\% of right-leaning outlets block access.

Additionally, we observe that the likelihood for a website to impose AI crawler restrictions decreases as its audience ideology diverges from the political center, with websites on both the far-left and far-right ends of the ideological spectrum imposing fewer AI crawler restrictions than moderate sources. Paralleling the increased restrictions on AI crawlers imposed by moderate sources, we observe their relative decline within training datasets in favor of hyperpartisan material. Through textual analysis of political content, we characterize the potential biases this shift may introduce, revealing over-representation of politically charged terminology in available training data.

These findings call for acute consideration in the curation of future training datasets to account for skewed restrictions in raw data collection, as the systematic withdrawal of high-quality news content, coupled with the relative  over-representation of politically hyperpartisan sources, demands active intervention by model developers to prevent embedding polarization into AI systems.

\section{Datasets}
In this article, we examine restrictions on AI crawlers deployed by the most popular websites worldwide. To this end, we rely on Google Chrome's analytics to identify popular websites and on Cloudflare Radar to categorize webpage content. Additionally, we combine a large-scale collection of \texttt{robots.txt} directives left by webmasters as of September 2025 and historical data since 2023 from CommonCrawl snapshots.

\paragraph{Websites Dataset (CrUX)} As a window into the web, we first consider in this study the set of the one million most visited websites worldwide accounting, in August 2025, for 95\% of global traffic on Google Chrome in terms of page loads \cite{Ruth2022WWW}. This dataset stems from the Chrome User Experience Report (CrUX) and represents aggregate analytics from hundreds of millions of Chrome users who have usage statistic reporting enabled, opted in to history syncing, and have no history sync passphrase set. Ruth et al. \cite{Ruth2022Top} performed comparative analysis of top lists, including the (deprecated) Alexa list and Cisco's Umbrella, and showcased the high coverage of the CrUX dataset. In addition to the mere list, we will leverage the popularity buckets disclosed in CrUX (top 1K, 10K, 100K, 1M). Unless stated otherwise, all analyses are performed over this set of websites or subsets thereof.

\paragraph{Content Categorization} We enrich this list of one million websites with a characterization of their content. To this end, we leverage, akin to Ruth et al \cite{Ruth2022WWW}, Cloudflare's Radar APIs \cite{Cloudflare_2025Domains}. Cloudflare operates, beyond DDoS protection, a DNS and parental control service whose filters allow parents or schools to block domains classified as ``Adult Themes'' or more generally falling under CIPA filters. To enable such filtering, Cloudflare characterized websites content ---thus extending beyond Cloudflare-protected websites---, we then rely on this categorization. For simplicity, we aggregate Cloudflare's granular content categories into ``supercategories'' ---we will drop the prefix for discussion fluidity--- relying on the taxonomy, reported in the Appendix, curated, and assessed in \cite{Ruth2022WWW}.

To validate Cloudflare's categorization, we extracted HTML meta keywords from 661k successfully crawled websites and identified overrepresented terms per category using chi-squared statistics. For instance, pages categorized as \textit{Shopping \& Auctions} tend to have keywords such as: \textit{silver, watches, jewellery}, while those categorized as \textit{Adult Themes} have: \textit{porn, telegram, sex}; we display the top words of each category in the Appendix.  

\paragraph{Robots.txt Collection} Additionally, we focus on the directives set by webmasters to crawlers scanning the web, indicating to them which portions of the website, if any, they are permitted to access. In September 2025, for each of the top one million websites we fetched the \texttt{robots.txt} file deployed, if any, at the root of their (sub)domains. Additionally, to go beyond a mere snapshot, we relied on CommonCrawl's periodic releases \cite{commoncrawl} of very-large-scale crawls of billions of webpages. In particular, this corpus contains timestamped \texttt{robots.txt} files, offering unparalleled chronological insights. However, rather than being a ``copy of the web'', only a fraction of it is crawled \cite{commoncrawl, Baack2024}. Within the top one million websites, on average over releases, 409k valid \texttt{robots.txt} files were collected. 

\section{Overview}

\subsection{Content Category}

We first aim to provide an overview of the web cross two dimensions: topic categories and languages, updating the description made by Ruth et al. \cite{Ruth2022WWW} in early 2022. We observe that half of the pages are written primarily in English (50.5\%), followed by Japanese (6.2\%) and Spanish (6.0\%), see figure in Appendix. Second, we observe in Figure \ref{fig:content_distrib} that \textit{Shopping \& Auctions} websites account for the largest fraction at 18.6\%. \textit{Business \& Economy}, \textit{Education}, \textit{Entertainment} and \textit{Technology} follow, representing between 7.6\% and 10.5\%. Aligned with \cite{Ruth2022WWW}, we observe that \textit{Adult} content shows significant variation across popularity tiers. Within the top thousand most visited websites worldwide, 20.3\% are \textit{Adult} websites, compared to 15.0\% in the top 10k and 3.7\% in the top 1M.

\begin{figure}[h]
    \centering
    \includegraphics[width=\columnwidth]{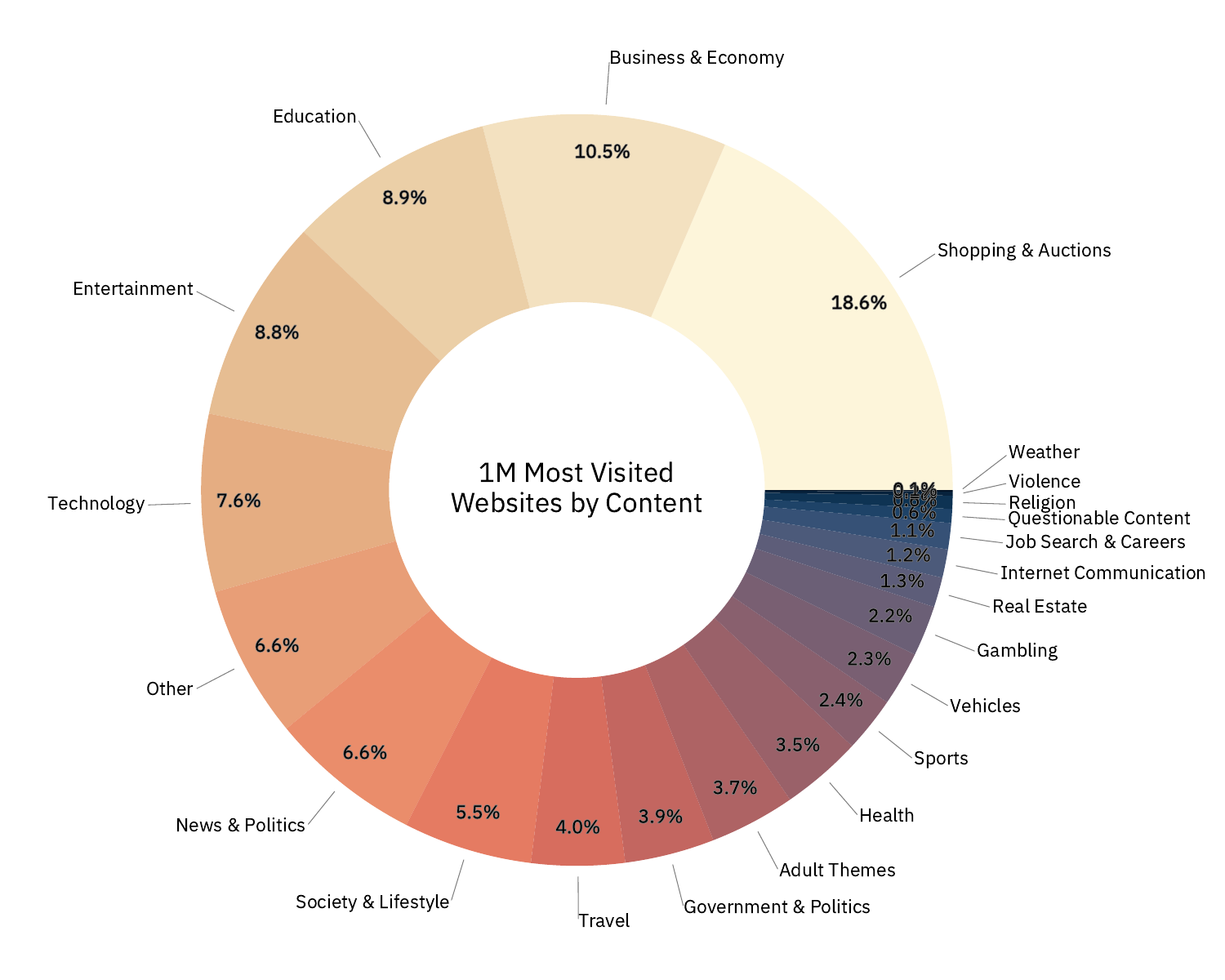}
   \caption{Distribution of content categories within the one million most visited websites worldwide (CrUX).}
    \label{fig:content_distrib}
\end{figure}

\subsection{Crawling Restrictions}

To prevent the content of their websites from being automatically indexed, scraped, and otherwise programmatically used without the website owner's agreement, multiple mechanisms have been developed to prevent such crawling.

In the top one million websites, we observe that 55.6\% of them deployed so-called \texttt{robots.txt} at the root of their (sub)domains.

Alternative proposals to prevent the crawling to feed AI model such as adding \texttt{noai} in the ``robots" meta tag of the page appeared on less than 0.2\% of analyzed websites, similarly \cite{dinzinger2024surveywebcontentcontrol} observed low adoption of the ``TDM Reservation Protocol''. In addition to such passive voluntary systems, Liu et al. \cite{2024arXiv241115091L} explored active measures whereby websites restrict the activity of some public crawlers, for instance by blocking traffic coming from IP addresses known to be operated by crawlers. Nevertheless, \texttt{robots.txt} is the \textit{de facto} standard to specify directives to web crawlers and as such will be further examined in the following section.

\section{Robots Exclusion Protocol}

With the advent of generative artificial intelligence, the datasets on which these models are trained have garnered attention and warranted renewed inspection of the Robots Exclusion Protocol \cite{hingle_knodel_2025_robotsTXT, dinzinger2024surveywebcontentcontrol, 2024arXiv241115091L}. We first consider a large-scale characterization of directives left by webmasters for crawlers before focusing specifically on those used by AI model providers.

\subsection{Overview}

The \texttt{robots.txt} can specify rules applying uniformly to all crawlers using the catch-all directive (Allow/Disallow for \texttt{User-Agent: *}) or by specifying targeted rules for particular crawlers; for instance, allowing Google Search's crawler to index certain pages while blocking all other robots. Within our collected sample, 64.1\% of \texttt{robots.txt} files used only the catch-all rules. 

Among crawlers named in \texttt{robots.txt} files of websites specifying granular directives, we can delineate the following categories:

\begin{itemize}
    \item \textbf{SEO \& Marketing}: \texttt{MJ12bot} (36.5\%), \texttt{AhrefsBot} (36.0\%), \texttt{AhrefsSiteAudit} (23.5\%), and \texttt{SemrushBot} (13.9\%)
    \item \textbf{Search Engines}: \texttt{Nutch} (25.1\%), \texttt{GoogleBot} (18.1\%), \texttt{CCBot} (14.7\%), \texttt{Yandex} (10.8\%), and \texttt{bingbot} (6.8\%)
    \item \textbf{Social Media}: \texttt{Pinterest} (23.6\%), \texttt{FacebookBot} (5.5\%), and \texttt{facebookexternalhit} (5.0\%)
    \item \textbf{Generative AI Training}: \texttt{GPTBot} (19.9\%), \texttt{ClaudeBot} (14.7\%), \texttt{Google-Extended} (14.4\%), \texttt{Applebot-Extended} (10.2\%), and \texttt{meta-externalagent} (8.8\%)
    \item \textbf{AI Search \& Chatbots}: \texttt{ChatGPT-User} (7.5\%), \texttt{PerplexityBot} (6.5\%), \texttt{Claude-Web} (5.2\%), \texttt{OAI-SearchBot} (4.9\%)
\end{itemize}

Alongside crawlers indexing the web for analytics, search engines, or generating social media previews, we observe AI-related ones. Granular distinctions emerge between crawlers collecting data for model training, e.g., \texttt{GPTBot, ClaudeBot}, those powering AI search engines, e.g., \texttt{OAI-SearchBot, PerplexityBot}, and those responding to individual user queries, e.g., \texttt{ChatGPT-User, Claude-Web}.

Importantly, the Robot Exclusion Protocol relies on good faith and voluntary compliance by the organizations operating crawlers. In 2025, Liu et al. \cite{2024arXiv241115091L} and Kim et al. \cite{kim2025} showed that AI crawlers operated by OpenAI, Anthropic, Meta, and CommonCrawl respected websites' robots.txt directives as stated. Yet, it should be noticed that the scope of these directives can be ambiguous. For instance, if a chatbot provider trains on user discussions without further curation, content from a website that blocked \texttt{GPTBot} but allowed (or, due to tacit consent, did not block) \texttt{ChatGPT-User} could still be used for training as it may appear in user chats.

\subsection{AI Crawlers}

\paragraph{2025 Snapshot} Focusing on the most popular AI chatbots \cite{CloudflareRobots}, we observe that, as of August/September 2025 and among the top one million websites, OpenAI's \texttt{GPTBot} is fully blocked on 10.6\% of websites, Anthropic's \texttt{ClaudeBot} on 9.1\%, and Google's \texttt{Google-Extended} on 8.9\%. Additionally, 9.5\% disallow \texttt{CCBot}, the crawler used by CommonCrawl, whose public datasets are extensively leveraged in training AI models \cite{gpt3,llama, Baack2024}. 

Alternatively, AI-powered search engine crawlers are more frequently permitted, with only 6.0\% of websites disallowing Perplexity's crawler, and 6.5\% and 6.1\% blocking OpenAI's and Anthropic's user-initiated crawlers (\texttt{ChatGPT-User} \& \texttt{Claude-Web}). As a baseline, 4.0\% of examined websites blocked Google Search's indexing crawler \texttt{GoogleBot} at their root.

We observed instances of websites, particularly news outlets, that disallowed popular AI crawlers except one due to licensing agreements \cite{openai2023axel-springer, openai2024newscorp}. Examples include the Wall Street Journal, New York Post, The Times, Bild, Washington Post, and Le Monde, all allowing OpenAI's crawlers while explicity disallowing others.

\begin{figure}[h]
    \centering
    \includegraphics[width=\columnwidth]{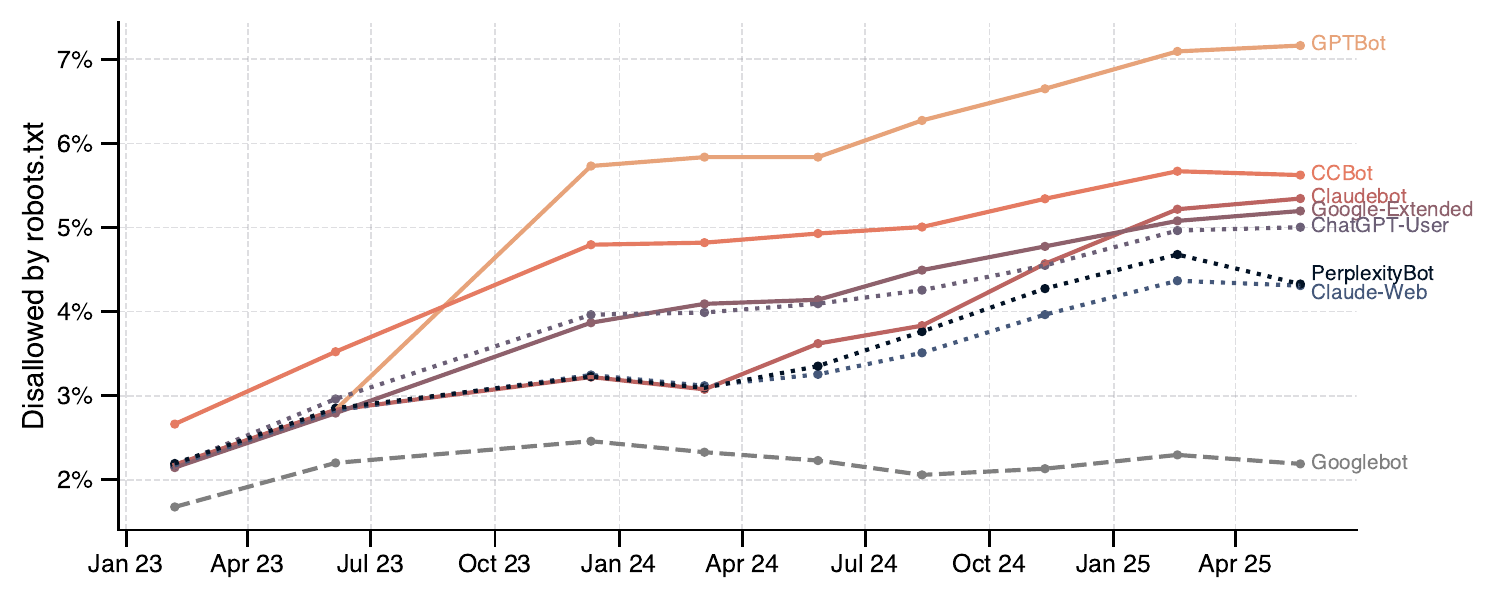}
    \caption{Fraction of webpages disallowing specific web crawlers on their domain over time, underlying data from CommonCrawl releases.}
    \label{fig:robots_txt_overtime}
\end{figure}

\paragraph{Chronology} Relying on CommonCrawl's periodic releases \cite{commoncrawl}, we can reconstruct the evolution of \texttt{robots.txt} since January 2023. We observe in Figure~\ref{fig:robots_txt_overtime} that while the fraction of webpages blocking Google Search's indexing crawler \texttt{GoogleBot} remains stable over the period 2023-2025, the fraction blocking OpenAI's \texttt{GPTBot} rises after its announcement in summer 2023 \cite{openaiCrawlers}. Following their public launch, other AI crawlers such as \texttt{ClaudeBot} and \texttt{PerplexityBot} experienced growing disallowance rates in \texttt{robots.txt} files throughout our observation window.

\begin{figure}[h]
    \centering
    \includegraphics[width=\columnwidth]{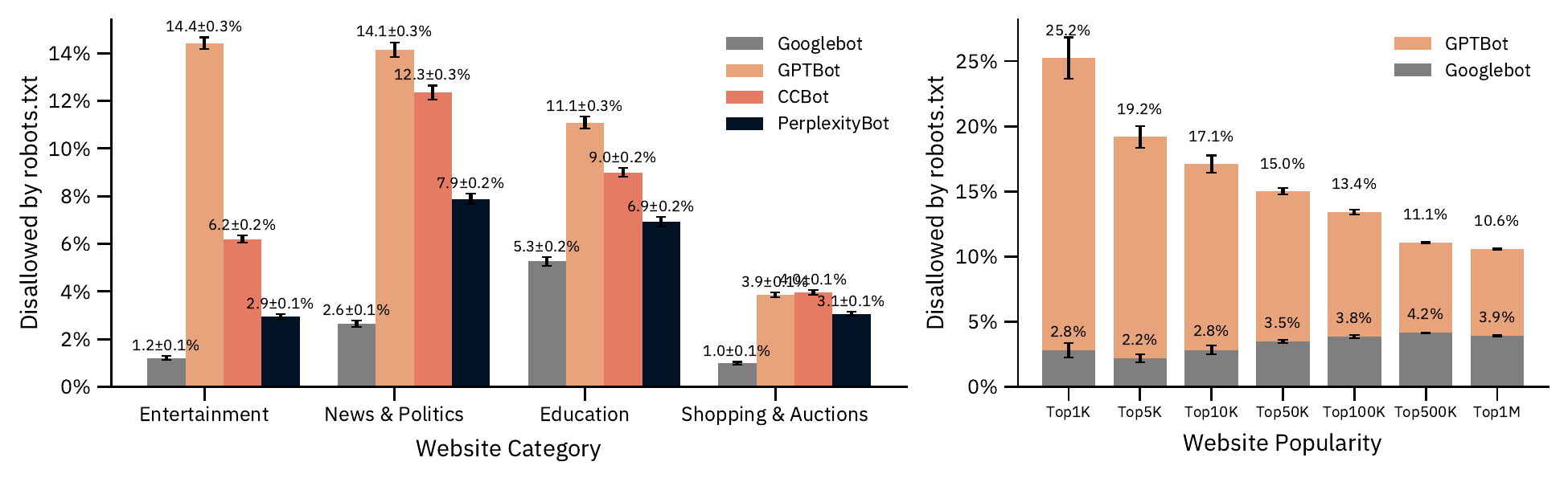}
    \caption{Fraction of webpages disallowing specific web crawlers on their domain, segmented by: (A) Content category per Cloudflare's classification, (B) CrUX popularity bucket. Error bars represent standard deviations over 100 bootstrap with replacement over websites.}
    \label{fig:robots_per_cat}
\end{figure}

\paragraph{Restrictions per category} Finally, we examine how robots exclusion directives are used across different categories of website content and popularity. Relying on Cloudflare categorization, Figure~\ref{fig:robots_per_cat} reveals that disallowance of OpenAI's \texttt{GPTBot} varies by topic across websites, with rates exceeding 14\% for Entertainment and News \& Politics, declining to 9.0\% for Education, and reaching only 4.0\% for Shopping \& Auctions. Additionally, we observe that more popular websites are more likely to disallow OpenAI's \texttt{GPTBot}: while 13.4\% of websites in the top 100k block it, this proportion nearly doubles to 25.2\% among the top thousand most visited websites worldwide.

\section{CommonCrawl \& Derivatives}

Since the observed directives that restrict crawlers from collecting training data for generative AI models are not uniformly distributed across the web, and because biases in training datasets can lead to discriminatory outputs in AI models \cite{basta-etal-2019-evaluating, modeltoobig, feng-etal-2023-pretraining}, we evaluate how the growing restriction of online resources impacts CommonCrawl and derivative datasets.

Indeed, refined datasets of CommonCrawl have emerged as fundamental in the development of recent large language models, representing over 80\% of training corpora of Llama \cite{llama} and GPT-3 \cite{gpt3}. We will in particular consider two subsets of CommonCrawl: FineWeb and FineWeb-Edu \cite{fineweb}. FineWeb is a cleaned and deduplicated English subset of CommonCrawl, while FineWeb-Edu is a subset of FineWeb containing English text with high ``educational quality'' \cite{fineweb}. Specifically, FineWeb-Edu was created by first scoring half a million FineWeb text snippets using a large language model, then by training a regression head over text embedding representations to extend this scoring to hundreds of millions of text snippets. Large language models trained on FineWeb-Edu were shown by HuggingFace in June 2024 to ``outperform all openly accessible web-based datasets on a number of reasoning- and knowledge-intensive benchmarks such as MMLU'' \cite{fineweb}.

\subsection{CommonCrawl}

First of all, to provide an overview of CommonCrawl's content, we consider a random sample of one billion tokens from its latest dump (\texttt{CC-MAIN-2025-38}) and analyze the language and content of the underlying crawled pages, relying on Cloudflare categorization \cite{Cloudflare_2025Domains}. Overall, aside from an under-representation of \textit{Adult} content compared to CrUX pages and an over-representation of pages in German, CommonCrawl appears analogous to the web in terms of language and content category; see the distribution in the Appendix.

\subsection{News Outlets}

\begin{figure}[h]
    \centering
    \includegraphics[width=\columnwidth]{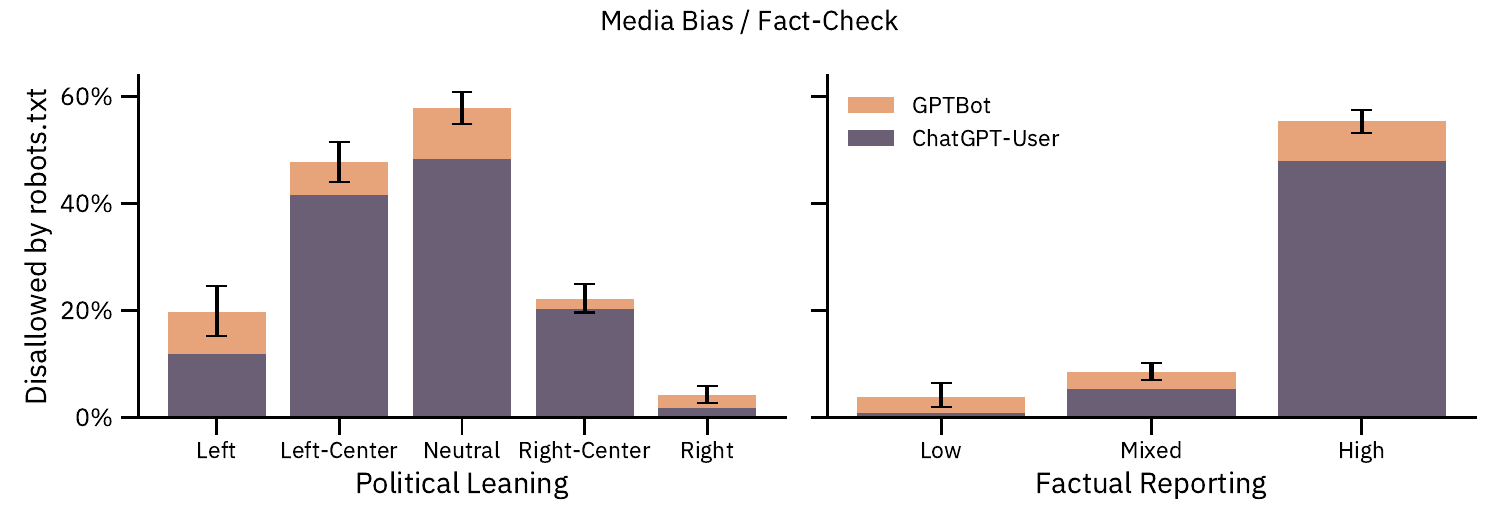}
    \caption{Fraction of websites disallowing OpenAI's \texttt{GPTBot} and \texttt{ChatGPT-User}, as a function of their political skew and factual reporting assessment by MBFC.}
    \label{fig:leaning_factual_block}
\end{figure}

Beyond this overall characterization, we now focus on news outlets. We relied on political skew and factual reporting assessments from Media Bias/Fact Check\footnote{https://mediabiasfactcheck.com}. We collected the \texttt{robots.txt} files of 3\,668 annotated news outlets. Overall, 34.2\% of media outlets disallowed OpenAI's \texttt{GPTBot}, with strong disparities across political leaning and factual reporting. As displayed in Figure \ref{fig:leaning_factual_block}, 58.0\% of \textit{neutral} outlets block OpenAI's \texttt{GPTBot}, compared to 19.6\% of \textit{left-leaning} outlets and 4.1\% of \textit{right-leaning} outlets. Regarding factual reporting, 55.4\% of outlets with \textit{high} factual reporting block OpenAI crawlers, compared to 8.4\% of those with \textit{mixed} factual reporting and 3.7\% of those with \textit{low} factual reporting. Additionally, we observe that across political leaning and factual reporting categories, the fraction of websites blocking training-related crawlers (\texttt{GPTBot}) is higher than those used to answer specific user requests (\texttt{ChatGPT-User}), as displayed in Figure \ref{fig:leaning_factual_block}.

In parallel, comparing the first Fineweb release from 2020 with that from 2025 (both larger than 170 billion tokens), the fraction of tokens stemming from sources evaluated as ``high factual reporting'' by MBFC decreased from 0.46\% to 0.26\% (a 41.3\% relative decrease). Similarly, the fraction of hyperpartisan right content increased from 19.3\% to 24.8\% (a 28.8\% relative increase), and of hyperpartisan left content rose from 18.9\% to 22.4\% (an 18.6\% relative increase).

\subsection{Educational Content}

Similarly, for educational resources, we examine FineWeb-Edu, a subset of FineWeb curated for its ``educational quality'' \cite{fineweb}. Importantly, each snippet in FineWeb-Edu is associated with the exact URL from which the content was extracted and its collection date by CommonCrawl's \texttt{CCBot}. This URL-level metadata enables us to assess \texttt{robots.txt} directives at the specific URL level rather than merely at the broader domain level.

For each release since 2015, we randomly sampled 25k URLs from domains in the top one million. We observe that over 20\% of the tokens in the FineWeb-Edu dataset collected prior to 2024 originated from domains that, as of August 2025, block CommonCrawl bots from accessing those very resources; see Figure in Appendix.

We preemptively seek to address the hypothesis that such restrictions on high-quality educational content may lead to overall decay in training dataset educational quality. For every FineWeb release since 2015, we randomly sample 10k text snippets and assess their ``educational quality'' using the language model used by HuggingFace to curate FineWeb-Edu \cite{fineweb}. The output is a score between 0 (not educational) and 5 (highly educational). We do not observe quality decay. Instead, the fraction of FineWeb (in tokens) scoring at least 3 (threshold to be included in FineWeb-Edu) grew from 8.3\% on average prior to 2022 to 9.7\% since 2022; see Figure in Appendix.

\subsection{Political Content}

Finally, we explore the directives left by webmasters to crawlers as a function of websites' political leaning. To assess such leaning, we rely on the large-scale audience characterization performed by Robertson et al. \cite{Robertson2018}. Specifically, they linked half a million Twitter accounts to US voter registration records and collected the URLs these accounts shared on the platform. The leaning of each domain was then assessed as the fraction of registered Democrats versus Republicans who had shared a URL from that domain, resulting in a continuous scale ranging from -1 to 1. A score of -1 (respectively +1) indicates that the domain was shared exclusively by Democrats (respectively Republicans), while a domain receives a score of 0 if and only if it was shared by equal proportions of Democrats and Republicans. This scale has been shown to align with other estimates derived from Facebook data, expert raters, and community assessments \cite{Robertson2018}.

\begin{figure}[h]
    \centering
    \includegraphics[width=\columnwidth]{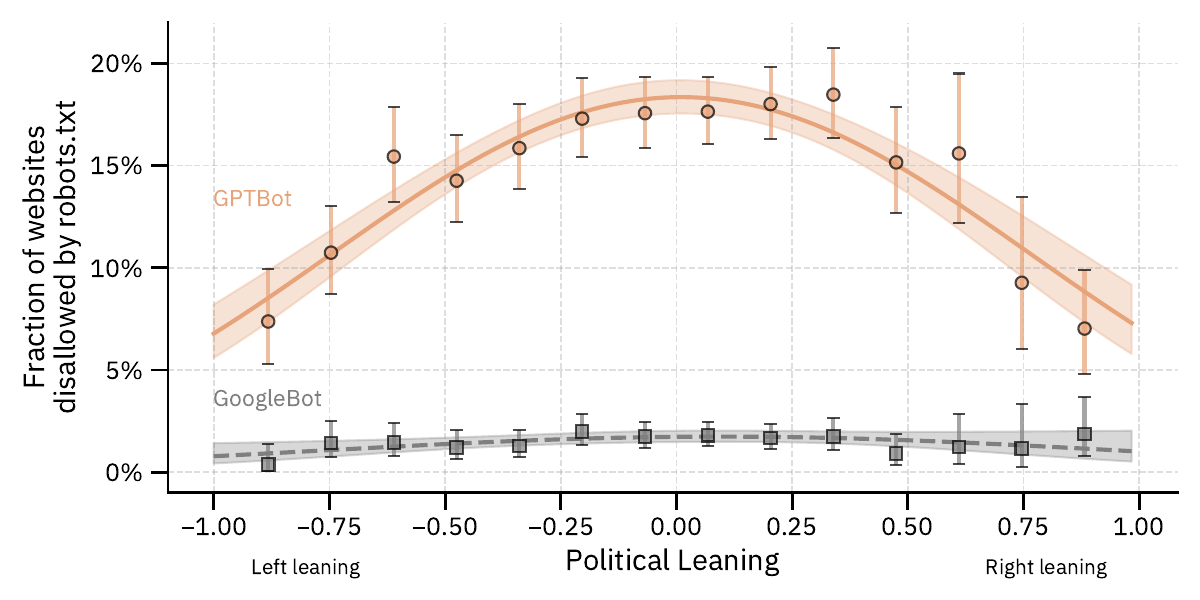}
    \caption{Fraction of websites disallowing \texttt{GPTBot} as a function of their audience ideological leaning, as characterized by Robertson et al \cite{Robertson2018}. Error bars represent Clopper-Pearson 95\% confidence intervals. The solid curve shows the fitted quadratic logistic regression with 95\% confidence band (shaded region). The fraction disallowing \texttt{GoogleBot} is shown as a baseline.}
    \label{fig:disallowed_by_robots_pol}
\end{figure}

\paragraph{Robot Exclusion} We collected the \texttt{robots.txt} files of 15.2k domains ideologically scaled by Robertson et al. \cite{Robertson2018} and display in Figure~\ref{fig:disallowed_by_robots_pol} the fraction of websites that disallow OpenAI's \texttt{GPTBot} as a function of their audience ideological leaning, with \texttt{GoogleBot} as a baseline. 

We observe that domains shared in similar proportions by Democrats and Republicans ---those with low absolute scores--- disallow OpenAI's \texttt{GPTBot} significantly more frequently than ideologically skewed domains. While 18.5\% of balanced domains (absolute score\footnote{Threshold of $\pm0.5$ set to align with AdFontes's labels.} below 0.5) block OpenAI's crawler, this fraction plummets to 7.3\% and 7.0\% for hyperpartisan left ($\leq -0.5$) and hyperpartisan right ($\geq 0.5$) domains, respectively. Similar results holds for CommonCrawl's \texttt{CCBot}. In contrast, in September 2022, the fraction of websites disallowing CommonCrawl's \texttt{CCBot} was equivalent between outlets with balanced and skewed audiences, at 2.6\% and 3.0\% respectively.

\paragraph{Representation in FineWeb}

Without inferring causation, we observe that the fraction of tokens extracted from hyperpartisan domains in FineWeb \cite{fineweb} increased from 27.7\% to 31.0\% between 2020 and 2025 (a 12.1\% relative increase).

To characterize potential biases that may arise from increased hyperpartisan content representation in FineWeb, we analyze word co-occurrence patterns in text authored by those websites. Our focus on co-occurrence rather than individual word frequencies aligns with large language model attention mechanisms, which are designed to capture contextual relationships between words. 

Specifically, from FineWeb latest release (\texttt{CC-MAIN-2025-26})  we filter text snippets from domains with characterized political leanings \cite{Robertson2018}. We split them into sentences and randomly sample five million sentences for each group: hyperpartisan left, hyperpartisan right, and balanced domains. Subsequently, we compute the co-occurrence within the same sentence of each word pair and apply chi-square statistics to identify word pairs that co-occur in the same sentence significantly more often in hyperpartisan text than in balanced content.

Among hyperpartisan left-leaning text, the following word pairs are among the most overrepresented compared to balanced text (arbitrary word order within pairs): \textit{(Climate; Change), (Human; Rights), (Grants; NIH), (Union; Workers), (Student; Yale), (Indigenous; Communities), (Women; Rights), (Women; Abortion), (People; Color), (Jewish; Community)}.

Similarly, among hyperpartisan right-leaning text: \textit{(President; Trump), (President; Biden), (AFA; Family), (Jesus; God), (Donald; Trump), (Christ; Jesus), (Fox; News), (Pro; Life), (Faith; God), (Supreme; Court)}.

For comparison purposes, we identify words that co-occur more frequently in hyperpartisan text alongside terms referring to potentially discriminated population:

\begin{itemize}
\item \textbf{Young}: (Left)\textit{ People, Women, Black}; (Right) \textit{God, Church, Men}
\item \textbf{Old}: (Left) \textit{Black, Scholar, Jewish}; (Right) \textit{Testament, God, Biden}
\item \textbf{Woman}: (Left) \textit{Black, White, Trans}; (Right) \textit{God, Abortion, Man}
\item \textbf{Man}: (Left) \textit{Black, White, Trump}; (Right) \textit{God, Jesus, Trump}
\item \textbf{Transgender}: (Left) \textit{People, LGBTQ, Rights}; (Right) \textit{Children, Women, School}
\item \textbf{LGBTQ}: (Left) \textit{Tennessee, Rhetoric, Safe}; (Right) \textit{Children, Agenda, Christian}
\item \textbf{Immigrant}: (Left) \textit{Republicans, Border, Detention}; (Right) \textit{Illegal, Border, Trump}
\item \textbf{Muslim}: (Left) \textit{Hindu, United, Americans}; (Right) \textit{Islamic, World, Brotherhood}
\end{itemize}

\section{Discussion}

The web's scale and diversity have established it as a foundational resource for training the AI models that power the development of foundation models. However, our analysis of the one million most visited websites worldwide reveals substantial and growing adoption of AI crawler restrictions, with over 25\% of the top thousand most visited websites blocking OpenAI's web crawler. Importantly, these restrictions are not uniformly distributed across the web; instead, they vary significantly by website popularity and content type, leading to potential skewing in subsequent web-crawled datasets. High-quality factual news outlets exhibit particularly pronounced restrictions, with over 55\% blocking of OpenAI's \texttt{GPTBot}.

Furthermore, we observe that outlets with high factual reporting and neutral political positions impose the strongest restrictions on OpenAI's crawlers, whereas hyperpartisan websites and outlets with low factual reporting impose fewer restrictions. Quantitatively, while 58\% of neutral outlets restrict OpenAI's GPTBot only 4.1\% of right-leaning outlets do so.

This pattern extends beyond news media to websites shared online more generally. Using Robertson et al.'s characterization of over 15k websites shared on Twitter \cite{Robertson2018}, we find that hyperpartisan websites---both left- and right-leaning---impose significantly fewer restrictions than politically balanced sources. While 18.5\% of balanced domains block OpenAI's GPTBot, only 7.3\% and 7.0\% of hyperpartisan left and right domains do so, respectively. Correspondingly, the representation of hyperpartisan domains in CommonCrawl-derived training datasets increased by 54.6\% (left) and 43.5\% (right) between pre-2023 and post-2023 periods, while balanced domains decreased by 11.3\%. Textual analysis reveals overrepresentation of politically charged, civic, and religious terminology in hyperpartisan content, with distinct semantic associations emerging for terms referring to marginalized groups.

These patterns raise concerns about systematic biases emerging in training datasets used to develop large language models. The withdrawal of high-quality news content and moderate political sources, combined with a relative increased representation of hyperpartisan material, may affect model outputs in ways that remain difficult to audit given the scale and opacity of training data compositions.

Importantly, our analysis focuses exclusively on \texttt{robots.txt} directives, the \textit{de facto} standard for programmatically communicating crawler restrictions. However, this represents only one form of access control. Longpre et al. \cite{ConsentInCrisis} demonstrated that when considering platform Terms of Service ---written in natural language rather than machine-readable format--- the fraction of C4, another large-scale CommonCrawl derivative, restricted from collection rose to 45\%, compared to only 5\% based on \texttt{robots.txt} alone. This suggests our findings may even underestimate the true extent of restricted content and the magnitude of potential biases introduced through selective data availability. Additionally, our focus on two English datasets, FineWeb \& FineWeb-Edu, while reflecting the emphasis given to English in AI model training \cite{llama, gpt3}, may overlook some biases, ignoring a potential education quality decay for non-English text for instance, and should be further investigated in the future.

The heterogeneous adoption of crawler restrictions, designed to protect content creators from exploitative practices, demands careful attention from AI model builders, who should undertake active curation strategies beyond mere boilerplate and heuristic-based content removal. Our work points to one potential approach: by establishing the distribution of crawling allowances, sampling strategies may be adjusted to preserve the representativeness of training datasets. Future research should examine how these compositional shifts affect downstream model performance, particularly regarding fairness, factual accuracy, and political neutrality in AI systems.

\bibliographystyle{ACM-Reference-Format}
\bibliography{sample-base}

\appendix
\section{Appendix}

\begin{table*}
\centering
\begin{tabular}{|l|p{7cm}|l|}
\hline
\textbf{Supercategory} & \textbf{Granular Categories} & \textbf{Top keywords} \\ \hline
Adult Themes & Pornography, Adult Themes, Nudity & porn, telegram, sex, free, videos \\ \hline
Business \& Economy & Business, Economy \& Finance, Brokerage \& Investing, Cryptocurrency, Professional Networking & smm, online, instagram, jobs, followers \\ \hline
Education & Educational Institutions, Education, Science, Space \& Astronomy, School Cheating & universitas, online, moodle, kampus, universit \\ \hline
News \& Politics & News \& Media, Government \& Politics & news, online, newspaper, latest, breaking \\ \hline
Government \& Politics & Politics, Advocacy, and Government-Related & online, government, development, portal, records \\ \hline
Entertainment & Audio Streaming, Music, Magazines, Cartoons \& Anime, Movies \& Home Video, Arts, Entertainment, Gaming, Video Streaming, Television, Comic Books, Paranormal & manga, anime, monster, sasaki, chapter \\ \hline
Gambling & Gambling & online, slot, result, casino, betting \\ \hline
Health & Health \& Fitness, Sex Education & health, medical, cosmetics, healthcare, care \\ \hline
Internet Communication & Forums, Webmail, Chat \& Messaging & forum, video, email, online, indian, free \\ \hline
Job Search \& Careers & Job Search \& Careers & jobs, academy, work, vacancies, career \\ \hline
Other & Redirect, Unknown & \\ \hline
Questionable Content & Drugs, Questionable Content, Questionable Activities, Alcohol, Hacking, Profanity & movies, free, watch, download, streaming \\ \hline
Real Estate & Real Estate & estate, casa, property, muebles, venta \\ \hline
Religion & Religion & suresi, prayer, bible, quran, church \\ \hline
Shopping \& Auctions & Ecommerce, Auctions \& Marketplaces, Coupons & silver, watches, jewellery, gold, necklace \\ \hline
Society \& Lifestyle & Lifestyle, Clothing and Fashion, Food \& Drink, Hobbies \& Interests, Home \& Garden, Pets, Parenting, Photography, Astrology, Dating \& Relationships, Arts \& Crafts, Sexuality, Tobacco, Body Art, Digital Postcards & dating, food, photo, baby, horoscope \\ \hline
Sports & Sports & football, live, soccer, sports, bike \\ \hline
Technology & Technology, File Sharing, Artificial Intelligence & free, download, apk, video, spotify \\ \hline
Travel & Travel &  flight, travel, bus, booking, hotel \\ \hline
Vehicles & Vehicles & duramax, car, diesel, auto, tuner \\ \hline
Violence & Weapons, Violence & holsters, gun, airsoft, firearms, reloading \\ \hline
Weather & Weather & weather, meteo, forecast, previsioni, italia \\ \hline
\end{tabular}
\caption{Taxonomy of Cloudflare's website content categorization. For each "super-category" we report some of the most overrepresented words in the meta tags of webpages categorized as such.}
\label{table:content_taxonomy}
\end{table*}

\begin{figure}
    \centering
    \includegraphics[width=\columnwidth]{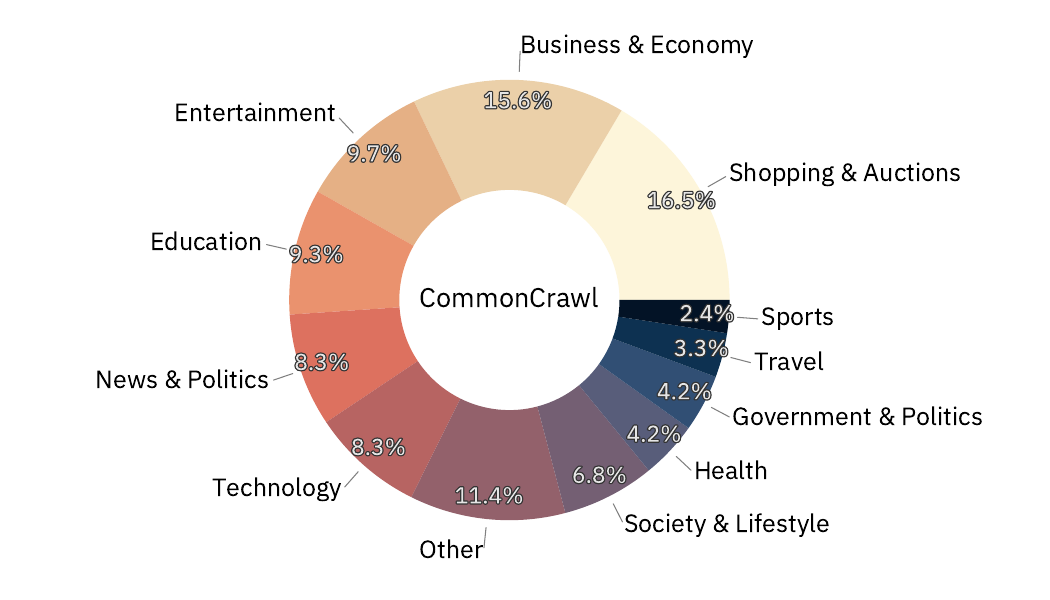}
    \caption{Distribution of topics in CommonCrawl, assessed through Cloudflare domain categorization. Categories less prevalent than 2\% are aggregated in ``other''.}
    \label{fig:cc_topic}
\end{figure}

\begin{figure}
    \centering
    \includegraphics[width=\columnwidth]{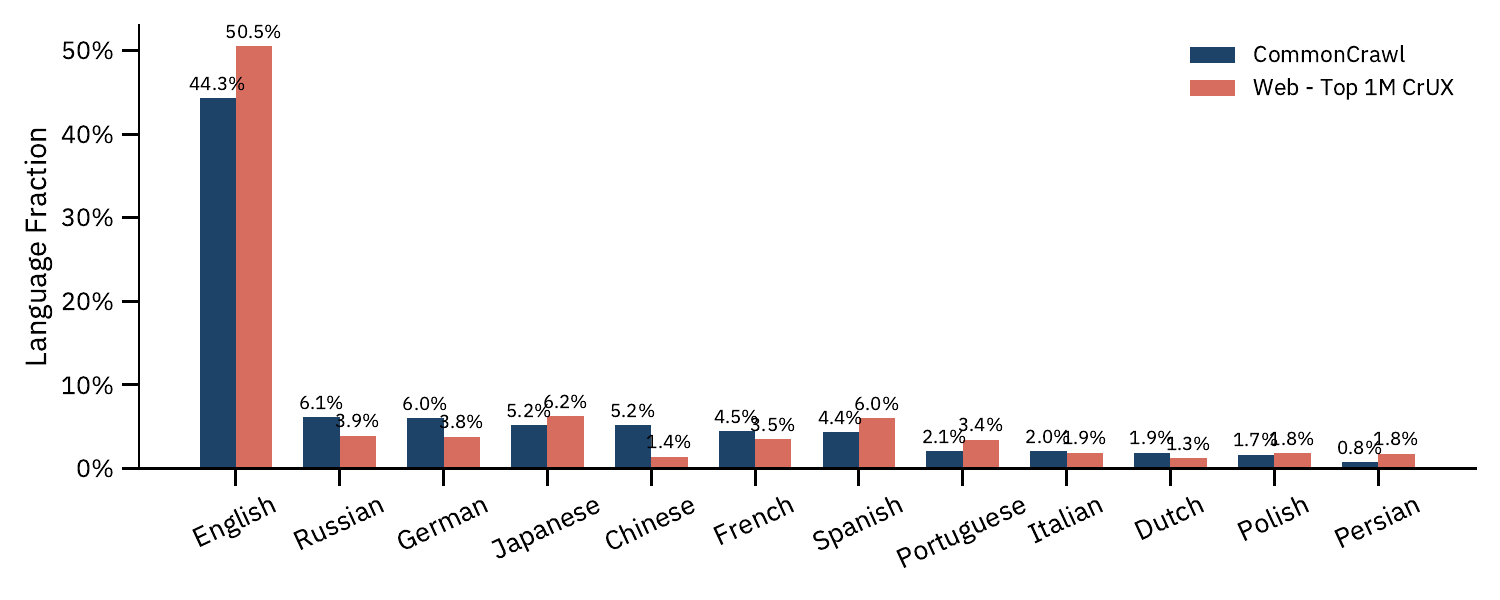}
    \caption{Distribution of languages of pages included in CommonCrawl and in the top one million in CrUX analytics. Importantly, because of a limited Google Chrome usage in People's Republic of China, the fraction of Chinese pages is abnormally low in CrUX analytics.}
    \label{fig:cc_crux_lang}
\end{figure}

\begin{figure*}[h]
    \centering
    \includegraphics[width=\textwidth]{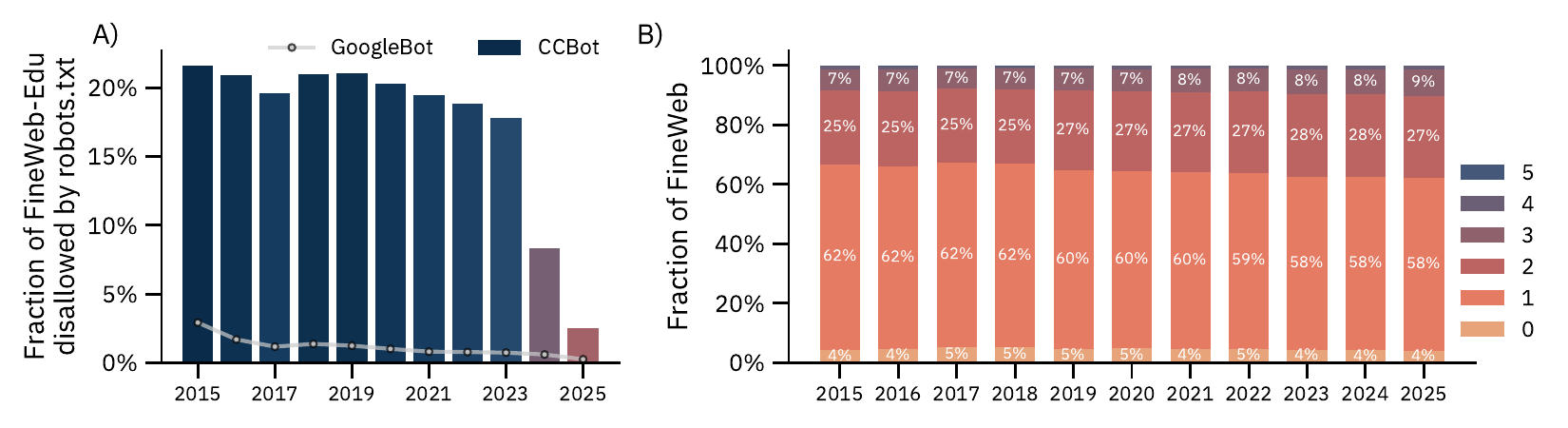}
    \caption{\textbf{(A)} Fraction (in tokens) of FineWeb-Edu sourced from websites that disallow \texttt{CCBot} as of August/September 2025, by year of collection. \textbf{(B)} Distribution of  ``educational quality'' score in FineWeb over time (text snippet score weighted by token count).}
    \label{fig:quality}
\end{figure*}

\end{document}